\documentclass[12pt,a4paper,reqno]{article}
\usepackage[letterpaper, left=2.5cm, right=2.5cm, top=1.5cm, bottom=1.5cm, headheight=7mm, headsep=0.8cm, footskip=1.5cm, includeheadfoot]{geometry}
\usepackage[utf8]{inputenc}
\usepackage{amsmath}
\usepackage{amssymb}
\usepackage{amsfonts}
\usepackage{stmaryrd}
\usepackage{bm}
\usepackage{amsthm}
\usepackage{mathtools}
\usepackage{algorithm}
\usepackage{algpseudocode}
\usepackage{color}
\usepackage{graphicx}
\usepackage{subcaption}
\usepackage[export]{adjustbox}
\usepackage{wrapfig}
\usepackage[numbers,square,compress]{natbib}
\usepackage[nottoc]{tocbibind}
\usepackage{tikz}
\usetikzlibrary{positioning}
\usepackage{hyperref}
\hypersetup{
    unicode=false,  
    pdftoolbar=true,
    pdfmenubar=true,
    pdffitwindow=false,
    pdfstartview={FitH},
    pdfnewwindow=true,
    colorlinks=true,
    linkcolor=blue,
    citecolor=blue,
    filecolor=blue,
    urlcolor=blue
}
\usepackage{authblk}
\usepackage{listings}

\newcommand{\revision}[1]{{\color{black}#1}}
\newcommand*{\doi}[1]{\href{http://dx.doi.org/\detokenize{#1}}{doi: \detokenize{#1}}}

\title{
{\bf Structure-preserving marker-particle discretizations of Coulomb collisions for particle-in-cell codes}}
\author[]{Eero Hirvijoki \thanks{eero.hirvijoki@gmail.com}}
\affil[]{Department of Applied Physics, Aalto University,\\ P.O. Box 11100, 00076 AALTO, Finland}

\date{
March 24, 2021}

\begin{document}
\maketitle
\begin{abstract}
This paper contributes new insights into discretizing Coulomb collisions in kinetic plasma models. Building on the previous works \cite{Carrillo-et-al:2020,Hirvijoki_Burby_metriplectic_gk}, I propose deterministic discrete-time energy- and positivity-preserving, entropy-dissipating marker-particle schemes for the standard Landau collision operator and the electrostatic gyrokinetic Landau operator. In case of the standard Landau operator, the scheme preserves also the discrete-time kinetic momentum. The improvements, the extensions of the structure-preserving discretizations in \cite{Carrillo-et-al:2020,Hirvijoki_Burby_metriplectic_gk} to discrete time, are made possible by exploiting the underlying metriplectic structure of the collision operators involved and the so-called discrete-gradient integrators.
\end{abstract}


\section{Introduction}
During the past ten years or so, numerical methods in kinetic plasma simulations have made quite a leap. We have seen the rise of structure-preserving, geometric particle-in-cell schemes \cite{Squire-Qin-Tang-PIC:2012PhPl,Evstatiev-shadwick:2013JCoPh,Shadwick-Stamm-Evstatiev:2014PhPl,Stamm-Shadwick-Evstatiev:2014ITPS,Xiao-et-al-kinetic:2015PhPl,He-et-al-Hamiltonian-splitting:2015PhPl,Qin-et-al:2016NucFu,Xiao-et-al-fluid:2016PhPl,Kraus-et-al:2017JPlPh,Xiao-et-al:2018PlST,Xiao-Qin-6d-tokamak:2020arXiv,Hirvijoki_Kormann_Zonta:2020PoP} that are based on discretizing either the variational or Hamiltonian structure of the underlying kinetic model. Such schemes are superior in that they preserve not only the energy of the system but typically guarantee also a local algebraic charge conservation law and the preservation of the multisymplectic two-form. These properties play an elementary role in targeting macroscopic transport time scales in kinetic simulations of fusion plasmas.

Beyond the dissipation-free Vlasov-Maxwell or Vlasov-Poisson part of kinetic models, there has been less development in obtaining structure-preserving discretizations for the dissipative part, namely the Coulomb collisions. This is likely due to the field simply being largely unaware of the underlying mathematical, so-called metriplectic structure of the Landau operator and, consequently, being unable to systematically exploit it the same way the variational and Hamiltonian structures have been exploited in developing structure-preserving discretizations for the dissipation-free part. It is true that metriplectic \cite{Kraus_Hirvijoki:2017PoP,Hirvijoki_Burby_Kraus:2018arXiv_galerkin} and other conservative discretizations \cite{Taitano_et_al:2015JCP,Hirvijoki_Adams:2017PhPl,Adams_et_al:2017SIAM,Hirvijoki_relativistic_landau:2019,Shiroto-Sentoku:2019PhRvE,Daniel_et_al:2020CPC,Shiroto_et_al:2020arXiv} exist for the standard Landau operator and even for its linearized version and model collision operators \cite{Crandall_et_al:2020CoPhC} but these schemes present the distribution function with a Galerkin-like basis or finite-differences methods. And, although attempts to combine the particle-based discretizations of the dissipation-free parts with Galerkin-like approaches for the dissipative part do exist \cite{Yoon-Chang:2014PoP,Hager_et_al:2016JCP,Hirvijoki_Kraus_Burby:2018arXiv_pic}, the major hurdle of preserving strict positivity of the distribution function remains. It is also unclear how the structure-preserving grid-based approaches would succeed in dealing with the rather complicated electrostatic gyrokinetic Landau operator which operates in the five-dimensional phase-space \cite{Burby_2015_collisions}.

It wasn't until very recently that the ideas of using marker-particle discretizations for the Landau operator truly started to pay off. While binary collision algorithms, pioneered by Takizuka and Abe \cite{takizukaabe:1977} and further \revision{studied by many \cite{Miller_Combi:1994,WANG1996,nanbu:1997,Nanbu_Yonemura:1998}}, have existed since 1977, such schemes are limited to using equal particle weights or otherwise \revision{do not guarantee energy and momentum conservation}. No algebraic proof of entropy dissipation exists either. Regarding this matter, we are now to thank Carrillo et al.~\cite{Carrillo-et-al:2020} for demonstrating how a deterministic particle-based discretization for the homogenous Landau collision operator, admitting arbitrary marker weights, can be constructed. Although \cite{Carrillo-et-al:2020} didn't present a scheme that would guarantee algebraic discrete-time energy conservation and entropy dissipation, continuous-time structure-preservation was obtained and the idea of interpreting Coulomb collisions as compressible flow driven by an entropy functional enabled a natural introduction of marker particles with arbitrary weights. This was a major improvement and lays the path forward to structure-preserving discrete-time formalism. 

In this paper, I will build on the work \cite{Carrillo-et-al:2020} and describe how to upgrade it to obtain the as-of-yet lacking discrete-time energy conservation and entropy dissipation for the standard Landau operator. Naturally, the scheme also provides discrete-time momentum conservation. Further, I will use the same technique to provide a discrete-time energy- and positivity-preserving, entropy-dissipating scheme for the electrostatic gyrokinetic Landau operator, extending the work \cite{Hirvijoki_Burby_metriplectic_gk}. \revision{During this process, it also becomes apparent how the quadratic expression for kinetic energy in particle coordinates would likely have admitted a discovery of the discrete-time energy conservation in the case of the standard Landau operator, given enough time and an educated guess. In the case of the significantly more involved electrostatic gyrokinetic Landau operator, however, I anticipate such discovery unlikely and the more systematic approach based on exploiting the metriplectic structure and the concept of discrete gradients to be not only convenient but necessary}. For the rest of the paper, in the order just described, I will discuss the standard Landau operator in Sec.~\ref{sec:regular-landau}, the electrostatic gyrokinetic version in Sec.~\ref{sec:gk-landau}, and provide a numerical demonstration of the conservation laws for the standard Landau operator in Sec.~\ref{sec:numerics}. I will not go through in detail the general framework of metriplectic dynamics. Instead, the reader is encouraged to consult \cite{Kaufman:1982fl,Kaufman:1984fb,Morrison:1984ca,Morrison:1984wu,Morrison:1986vw,Grmela:1984dn,Grmela:1984ea,Grmela:1985jd} for more details and examples in that topic.

Finally, although these latest developments are rather encouraging, yet new questions arise. There is, e.g., no proof yet that the variational integrators or Hamiltonian splitting schemes used in the modern structure-preserving particle-in-cell methods would guarantee the conservation of the very entropy functional the particle-based collision operators now rely upon and are to dissipate. \revision{Also, since symplectic integrators do not exactly preserve the system energy but, instead, provide a strict bound for it to fluctuate within, investigating the coupling of these algorithms would be important.} Analysis of such question is left for future studies, though.

\section{The standard Landau collision operator}\label{sec:regular-landau}
This section is concerned with communicating two results: (i) presenting the discrete-time energy- and momentum-conserving, entropy-dissipating scheme for the standard Landau operator and (ii) explaining how the result naturally emerges when considering the associated metriplectic structure and discrete gradients. 
I will use the notation $\bm{z}=(\bm{x},\bm{v})$ to denote the phase-space coordinates. To distinguish between two different phase-space locations, I will introduce also the notation $\bm{\bar{z}}=(\bm{\bar{x}},\bm{\bar{v}})$. 
\subsection{The Vlasov-Maxwell-Landau model}
The Vlasov-Maxwell-Landau model is a reasonable kinetic description for many hot plasmas. In short, the system can be described with the dynamic equations
\begin{align}
    \frac{\partial f_s}{\partial t}+\nabla\cdot (\bm{v}f_s)+\frac{\partial}{\partial \bm{v}}\cdot\left[\frac{e_s}{m_s}\left(\bm{E}+\bm{v}\times\frac{\bm{B}}{c}\right)\revision{f_s}\right]&=\sum_{\bar{s}}C_{s\bar{s}}(f_s,f_{\bar{s}}),\\
    \frac{1}{c}\frac{\partial \bm{E}}{\partial t}+\frac{4\pi}{c} \sum_se_s\int \bm{v}f_sd\bm{v}&=\nabla\times\bm{B},\\
    \frac{1}{c}\frac{\partial \bm{B}}{\partial t}+\nabla\times\bm{E}&=0,
\end{align}
that are supplemented with the constraining, static equations
\begin{align}
    \nabla\cdot\bm{E}&=4\pi \sum_se_s\int f_sd\bm{v},\\
    \nabla\cdot\bm{B}&=0.
\end{align}
Often in kinetic simulations, the collision operator is either neglected, on the basis of studying only short time scales, or approximated. Here, I will use the full form of it, though.

The Landau operator, describing the effects due to small-angle Coulomb collisions between the species $s$ and $\bar{s}$, which can be the same species as well, reads
\begin{align}
    C_{s\bar{s}}(f_s,f_{\bar{s}})=-\frac{\nu_{s\bar{s}}}{m_s}\frac{\partial}{\partial \bm{v}}\cdot\bm{\gamma}_{s\bar{s}},
\end{align}
where $\nu_{s\bar{s}}=2\pi e_se_{\bar{s}}\ln\Lambda$ and the three-component vector $\bm{\gamma}_{s\bar{s}}$ is given by
\begin{align}
    \bm{\gamma}_{s\bar{s}}=\int \delta(\bm{x}-\bm{\bar{x}})f_s(\bm{z})f_{\bar{s}}(\bm{\bar{z}})\mathbb{Q}(\bm{v}-\bm{\bar{v}})\cdot\bm{\Gamma}_{s\bar{s}}(\mathcal{S},\bm{z},\bm{\bar{z}})d\bm{\bar{z}}.
\end{align}
The expression for $\bm{\gamma}_{s\bar{s}}$ differs from the one typically used in the literature for I have re-organized it somewhat to explicitly introduce the entropy functional 
\begin{align}
    \label{eq:fk-entropy-functional}
    \mathcal{S}=-\sum_s\int f_s\ln f_s d\bm{z}.
\end{align}
In the expression for $\bm{\gamma}_{s\bar{s}}$, the vector $\bm{\Gamma}_{s\bar{s}}(\mathcal{A},\bm{z},\bm{\bar{z}})$ is given by
\begin{align}
    \bm{\Gamma}_{s\bar{s}}(\mathcal{A},\bm{z},\bm{\bar{z}})=\frac{1}{m_s}\frac{\partial}{\partial \bm{v}}\frac{\delta\mathcal{A}}{\delta f_s}(\bm{z})
    -\frac{1}{m_{\bar{s}}}\frac{\partial}{\partial \bm{\bar{v}}}\frac{\delta\mathcal{A}}{\delta f_{\bar{s}}}(\bm{\bar{z}}),
\end{align}
the matrix $\mathbb{Q}(\bm{\xi})$ is a scaled projection onto a plane perpendicular to $\bm{\xi}$
\begin{align}
    \mathbb{Q}(\bm{\xi})=\frac{1}{|\bm{\xi}|}\left(\mathbb{I}-\frac{\bm{\xi}\bm{\xi}}{|\bm{\xi}|^2}\right),
\end{align}
with $\mathbb{I}$ the identity matrix in three dimensions, 
and the functional derivative is defined via the Fr\'echet derivative
\begin{align}
    \frac{\partial}{\partial \epsilon}\Big|_{\epsilon=0}\mathcal{A}[f_s+\epsilon\delta f_s]=\int \frac{\delta \mathcal{A}}{\delta f_s}\delta f_sd\bm{z}\equiv \delta\mathcal{A}[\delta f_s].
\end{align}
\revision{The standard form of the collision operator is recovered after computing the functional derivative of the entropy 
\begin{align}
    \frac{\delta \mathcal{S}}{\delta f_s}=-1-\ln f_s,
\end{align}
and then evaluating the expression 
\begin{align}
    f_s(\bm{z})f_{\bar{s}}(\bar{\bm{z}})\bm{\Gamma}_{s\bar{s}}(\mathcal{S},\bm{z},\bm{\bar{z}})=\frac{f_{s}(\bm{z})}{m_{\bar{s}}}\frac{\partial f_{\bar{s}}}{\partial \bm{\bar{v}}}-\frac{f_{\bar{s}}(\bar{\bm{z}})}{m_s}\frac{\partial f_{s}}{\partial \bm{v}}.
\end{align}
}

The Landau operator conserves the kinetic energy functional
\begin{align}
    \mathcal{K}=\sum_s\int\frac{1}{2}m_s|\bm{v}|^2f_sd\bm{z},
\end{align}
in a local sense. That is, in collisions between any two species $s$ and $\bar{s}$, the collision operator satisfies the relation
\begin{align}
    \int \delta(\bm{x}-\bm{x}')\frac{1}{2}m_s|\bm{v}|^2C_{s\bar{s}}(f_s,f_{\bar{s}})d\bm{z}+\int \delta(\bm{\bar{x}}-\bm{x}')\frac{1}{2}m_{\bar{s}}|\bm{\bar{v}}|^2C_{\bar{s}s}(f_{\bar{s}},f_s)d\bm{\bar{z}}=0
\end{align}
for any choice of the location $\bm{x}'$. Similarly the collision operator conserves the kinetic momentum functional
\begin{align}
    \bm{\mathcal{P}}=\sum_s\int m_s\bm{v}f_sd\bm{z},
\end{align}
in a local sense. That is, the Landau operator satisfies
\begin{align}
    \int \delta(\bm{x}-\bm{x}')m_s\bm{v}C_{s\bar{s}}(f_s,f_{\bar{s}})d\bm{z}+\int \delta(\bm{\bar{x}}-\bm{x}')m_{\bar{s}}\bm{\bar{v}}C_{\bar{s}s}(f_{\bar{s}},f_s)d\bm{\bar{z}}=0,
\end{align}
again for arbitrary $\bm{x}'$. In short, at every configuration space position, the kinetic energy and momentum transfer equally between two different species during collisions and not at all in self collisions.

\subsection{Metric formulation of the collision operator}
The choice of expressing the Landau operator in terms of an entropy functional is an intentional one. 
Where the dissipation-free Vlasov-Maxwell dynamics are well known to have an infinite-dimensional non-canonical Hamiltonian structure culminating into a functional Poisson bracket and driven by a Hamiltonian functional \cite{Morrison:1980PhLA,Weinstein_Morrison:1981,Marsden_Weinstein:1982PhyD}, the collisional dynamics can be formulated as an infinite-dimensional metric bracket and driven by the entropy functional. Together the two brackets form a so-called metriplectic structure for the Vlasov-Maxwell-Landau system. A detailed account on this matter is given in \cite{Morrison:1986vw,Kraus_Hirvijoki:2017PoP} and will not be repeated here. Instead, I outline the derivation of the metric bracket only to motivate the later use of it in the discretizations.

Generalizing the study of collisional dynamics to arbitrary functions, I multiply the collision operators of different species with arbitrary species-dependent functions $G_s(\bm{z})$. Then I integrate the expressions over the associated phase spaces and sum everything together. The result is
\begin{align}
    \label{eq:fk-collisional-evolution}
    &\sum_{s\revision{,\bar{s}}}\int G_s(\bm{z})C_{s\bar{s}}(f_s,f_{\bar{s}})d\bm{z}\nonumber\\
    &=\sum_{s,\bar{s}}\iint\frac{1}{m_s}\frac{\partial G_s(\bm{z})}{\partial\bm{v}}\cdot\mathbb{W}_{s\bar{s}}(\bm{z},\bm{\bar{z}})\cdot\bm{\Gamma}_{s\bar{s}}(\mathcal{S},\bm{z},\bm{\bar{z}})d\bm{\bar{z}}d\bm{z}\nonumber\\
    &=\sum_{s,\bar{s}}\frac{1}{2}\iint\left(\frac{1}{m_s}\frac{\partial G_s(\bm{z})}{\partial\bm{v}}-\frac{1}{m_{\bar{s}}}\frac{\partial G_{\bar{s}}(\bm{\bar{z}})}{\partial\bm{\bar{v}}}\right)\cdot\mathbb{W}_{s\bar{s}}(\bm{z},\bm{\bar{z}})\cdot\bm{\Gamma}_{s\bar{s}}(\mathcal{S},\bm{z},\bm{\bar{z}})d\bm{\bar{z}}d\bm{z},
\end{align}
where the positive semidefinite matrix $\mathbb{W}_{s\bar{s}}(\bm{z},\bm{\bar{z}})$ is defined as
\begin{align}
    \label{eq:fk-W-matrix}
    \mathbb{W}_{s\bar{s}}(\bm{z},\bm{\bar{z}})=\nu_{s\bar{s}} \delta(\bm{x}-\bm{\bar{x}})f_s(\bm{z})f_{\bar{s}}(\bm{\bar{z}})\mathbb{Q}(\bm{v}-\bm{\bar{v}}),
\end{align}
and the last line follows from the antisymmetry of the vector $\bm{\Gamma}_{s\bar{s}}(\mathcal{A},\bm{z},\bm{\bar{z}})$ with respect to the positions and the species labels.

If I now consider the function $G_s$ to be expressed as a functional derivative instead
\begin{align}
    \mathcal{G}=\sum_s\int G_sf_sd\bm{z} \qquad \implies \qquad \frac{\delta \mathcal{G}}{\delta f_s}=G_s,
\end{align}
I can identify $\bm{\Gamma}_{s\bar{s}}(\mathcal{G},\bm{z},\bm{\bar{z}})$ in the expression \eqref{eq:fk-collisional-evolution} and define a symmetric, positive semidefinite bracket 
\begin{align}\label{eq:fk-metric-bracket}
    (\mathcal{A},\mathcal{B})=\sum_{s,\bar{s}}\frac{1}{2}\iint\bm{\Gamma}_{s\bar{s}}(\mathcal{A},\bm{z},\bm{\bar{z}})\cdot\mathbb{W}_{s\bar{s}}(\bm{z},\bm{\bar{z}})\cdot\bm{\Gamma}_{s\bar{s}}(\mathcal{B},\bm{z},\bm{\bar{z}})d\bm{\bar{z}}d\bm{z}.
\end{align}
In terms of this bracket, the collisional evolution of arbitrary functionals can be generalized to the functional differential equation
\begin{align}
    \frac{d\mathcal{A}}{dt}\Big|_{\text{coll}}=(\mathcal{A},\mathcal{S}).
\end{align}
It is then straightforward to verify that the kinetic energy and momentum functionals are invariants of the bracket in the sense that $(\mathcal{K},\mathcal{A})=0$ and $(\bm{\mathcal{P}},\mathcal{A})=\bm{0}$ with respect to arbitrary $\mathcal{A}$ and consequently the collisional evolution of these functionals vanishes. Furthermore, it is straightforward to verify that evaluating the bracket with respect to the functional $\int \delta(\bm{x}-\bm{x}')\delta(\bm{v}-\bm{v}')f_s(\bm{z}')d\bm{z}'$ and the entropy \eqref{eq:fk-entropy-functional} results in the collision operator.

\subsection{Marker-particle discretization of the collisional bracket}
\revision{For the rest of this section, only the single-species case is considered to avoid clutter of indices. Consequently, the species indices $s$, $\bar{s}$ and the associated summations are dropped. All of the presented results regarding the discrete conservation properties propagate to the multiple-species case as well, although this is not explicitly demonstrated.}

I will now consider a discretization common in particle-in-cell simulations, namely that the distributional density is presented as
\begin{align}\label{eq:marker-distribution}
    f_h(\bm{z},t)d\bm{z}=\sum_pw_p\delta(\bm{x}-\bm{x}_p(t))\delta(\bm{v}-\bm{v}_p(t))d\bm{z}.
\end{align}
Before such particle-based discretization can be seriously considered, though, a couple of obvious questions need to be addressed. 

The first question relates to evaluating the entropy functional. If evaluated with respect to the marker-particle distribution \eqref{eq:marker-distribution}, the entropy functional \eqref{eq:fk-entropy-functional} is not well defined. A resolution to this issue was proposed and discussed in great detail in \cite{Carrillo-et-al:2020}. Introducing a regularized entropy functional, where the distribution function is first convoluted with a radial basis function $\psi_\epsilon$, the entropy functional can be replaced with the functional 
\begin{align}
    \mathcal{S}[f]\rightarrow \mathcal{S}[\psi_\epsilon\ast f]\equiv \mathcal{S}_{\epsilon}[f].
\end{align}
This enables the evaluation of $\mathcal{S}_{\epsilon}[f]$ even for the delta distribution \eqref{eq:marker-distribution}. In practice, one can simply replace the delta functions with radial basis functions in evaluating the regularized entropy functional so that
\begin{align}
    \label{eq:fk-regularized-entropy}
    \mathcal{S}[\psi_\epsilon\ast f_h]=-\int \sum_pw_p\psi_\epsilon(\bm{z}-\bm{z}_p)\ln\left(\sum_{p'}w_{p'}\psi_\epsilon(\bm{z}-\bm{z}_{p'})\right)d\bm{z}.
\end{align} 

The second question relates to evaluating the functional derivatives with respect to the distribution \eqref{eq:marker-distribution}. A solution to this issue, too, was proposed in Ref.~\cite{Carrillo-et-al:2020}. Here, I approach this second matter from a slightly different perspective, though.
Recall the variation of a functional $\mathcal{A}[f]$ as defined via the Fr\'echet derivative
\begin{align}
    \delta \mathcal{A}[\delta f]=\int \frac{\delta \mathcal{A}}{\delta f}\delta f d\bm{z}.
\end{align}
If the function $f$ is restricted to the specific parametrized form of $f_h$ in \eqref{eq:marker-distribution}, also the perturbation becomes parametrized and  takes the following specific form
\begin{align}
    \delta f_hd\bm{z}=-\sum_p w_p\left(\delta(\bm{v}-\bm{v}_p)\nabla\delta(\bm{x}-\bm{x}_p)\cdot\delta \bm{x}_p+\delta(\bm{x}-\bm{x}_p)\frac{\partial}{\partial \bm{v}}\delta(\bm{v}-\bm{v}_p)\cdot\delta \bm{v}_p\right)d\bm{z}.
\end{align}
Consequently, after partial integrations, the variation of the functional $\mathcal{A}[f]$, restricted to the distributions of type \eqref{eq:marker-distribution}, becomes 
\begin{align}
    \delta \mathcal{A}[\delta f_h]
    &=\sum_pw_p\left(\nabla\frac{\delta \mathcal{A}}{\delta f}\Big|_{\bm{z}_p}\cdot\delta\bm{x}_p+\frac{\partial }{\partial \bm{v}}\frac{\delta \mathcal{A}}{\delta f}\Big|_{\bm{z}_p}\cdot \delta \bm{v}_p\right).
\end{align}
On the other hand, when the functional $\mathcal{A}$ is evaluated with respect to the distribution $f_h$, it becomes an ordinary function of the degrees of freedom, namely 
\begin{align}
    \mathcal{A}[f_h]=A(\bm{X},\bm{V};W),
\end{align}
with $\bm{X}=\{\bm{x}_p\}_p$ and $\bm{V}=\{\bm{v}_p\}_p$ denoting the collections of the marker-particle degrees of freedoms and $W=\{w_p\}_p$ the collection of particle weights that are treated as fixed parameters. Consequently, the variation $\delta\mathcal{A}[\delta f_h]$ has to be equal to the variation of the function $A(\bm{X},\bm{V};W)$, which provides the identities
\begin{align}
    \label{eq:discrete_functional_spatial_derivative}
    \nabla\frac{\delta \mathcal{A}}{\delta f}\Big|_{\bm{z}_p}&=\frac{1}{w_p}\frac{\partial A(\bm{X},\bm{V};W)}{\partial \bm{x}_p},\\
    \label{eq:discrete_functional_velocity_derivative}
    \frac{\partial}{\partial \bm{v}}\frac{\delta \mathcal{A}}{\delta f}\Big|_{\bm{z}_p}&=\frac{1}{w_p}\frac{\partial A(\bm{X},\bm{V};W)}{\partial \bm{v}_p}.
\end{align}

The final detail that needs to be resolved before the collisional bracket can be evaluated with respect to the discrete distribution is the local nature of the Landau operator. In numerical implementations, two marker particles in a particle-in-cell code will, in practice, never be at exactly the same location. Consequently the delta function in \eqref{eq:fk-W-matrix}, responsible for this localization, has to be approximated. I suggest dividing the spatial configuration domain into disjoint, so-called collision cells that share a boundary if adjacent. This is a standard practise in numerical implementations relying on the binary collision algorithms. In such case, the delta function can be replaced by an indicator function $\mathbf{1}(p,\bar{p})$ which is one if the particles $p$ and $\bar{p}$ are within the same collision cell and zero otherwise.

Substituting the discrete distribution function \eqref{eq:marker-distribution}, the transformation of the functional derivatives \eqref{eq:discrete_functional_velocity_derivative}, and the replacement of the localizing delta function with an indicator function into the single-species version of the metric bracket \eqref{eq:fk-metric-bracket}, I recover a finite-dimensional bracket 
\begin{align}
    \label{eq:fk-bracket-discrete}
    (A,B)_h=\frac{1}{2}\sum_{p,\bar{p}}\bm{\Gamma}(A,p,\bar{p})\cdot\mathbb{W}(p,\bar{p})\cdot\bm{\Gamma}(B,p,\bar{p}),
\end{align}
where the vector $\bm{\Gamma}(A,p,\bar{p})$ is defined as
\begin{align}
    \bm{\Gamma}(A,p,\bar{p})=\frac{1}{mw_p}\frac{\partial A}{\partial \bm{v}_p}-\frac{1}{mw_{\bar{p}}}\frac{\partial A}{\partial \bm{v}_{\bar{p}}},
\end{align}
and the matrix $\mathbb{W}(p,\bar{p})$ is given by
\begin{align}
    \mathbb{W}(p,\bar{p})=\nu w_pw_{\bar{p}}\mathbf{1}(p,\bar{p})\mathbb{Q}(\bm{v}_p-\bm{v}_{\bar{p}}).
\end{align}

The finite-dimensional bracket \eqref{eq:fk-bracket-discrete} provides collisional evolution of functions that depend on the particle degrees of freedom via the equation
\begin{align}
    \label{eq:fk-ode}
    \frac{d A}{dt}\Big|_{\text{coll}}=(A,S_\epsilon)_h,
\end{align}
with $S_\epsilon(\bm{X},\bm{V};W)=\mathcal{S}_\epsilon[f_h]$ the regularized entropy functional defined in \eqref{eq:fk-regularized-entropy}. Of specific interest is the case $A=\bm{v}_p$ which provides the equation of motion for the particle coordinates. Exploiting the antisymmetry of the vector $\bm{\Gamma}(A,p,\bar{p})$ with respect to exchanging $p$ and $\bar{p}$, the result can be expressed as
\begin{align}
    \label{eq:fk-vdot}
    \frac{d\bm{v}_p}{dt}\Big|_{\text{coll}}=\frac{\nu}{m}\sum_{\bar{p}}w_{\bar{p}} \mathbf{1}(p,\bar{p})\mathbb{Q}(\bm{v}_p-\bm{v}_{\bar{p}})\cdot\bm{\Gamma}(S_\epsilon,p,\bar{p}).
\end{align}
Apart from the indicator function, which arises from considering also the configuration space coordinates $\bm{x}_p$, and the different reasoning of the discrete functional derivative, this result is identical to the one derived in \cite{Carrillo-et-al:2020} with different argumentation.

The finite-dimensional bracket \eqref{eq:fk-bracket-discrete} preserves the discrete kinetic energy
\begin{align}
    K(\bm{X},\bm{V};W)&=\mathcal{K}[f_h]=\sum_pw_p\frac{m}{2}|\bm{v}_p|^2,\\
    \frac{d K}{dt}\Big|_{\text{coll}}&=(K,S_\epsilon)_h=0.
\end{align}
This follows from the fact that the discrete kinetic energy is an invariant of the discrete bracket in the sense of $(K,A)_h=0$ with respect to arbitrary functions $A$. This is straightforward to verify after realizing that 
    $\bm{\Gamma}(K,p,\bar{p})=\bm{v}_p-\bm{v}_{\bar{p}}$,
and, consequently, $\bm{\Gamma}(K,p,\bar{p})\cdot\mathbb{W}(p,\bar{p})=0$. In addition to the kinetic energy, also the discrete kinetic momentum 
\begin{align}
    \bm{P}(\bm{X},\bm{V};W)&=\bm{\mathcal{P}}[f_h]=\sum_pw_pm\bm{v}_p,\\
    \frac{d\bm{P}}{dt}\Big|_{\text{coll}}&=(\bm{P},S_\epsilon)_h=0,
\end{align}
is conserved. This, too, follows from the fact that the discrete kinetic momentum is an invariant of the discrete bracket, i.e., $(\bm{P},A)_h=0$ with respect to arbitrary $A$, which follows trivially from the condition
    $\bm{\Gamma}(\bm{P},p,\bar{p})=\mathbb{I}-\mathbb{I}=0$.
Finally, as the discrete bracket is positive semidefinite, the regularized entropy is trivially dissipated 
\begin{align}
    \frac{dS_\epsilon}{dt}\Big|_{\text{coll}}&=(S_\epsilon,S_\epsilon)_h\ge 0.
\end{align}

\subsection{A simple conservative integrator}
I will first introduce a simple modification of the temporal discretization used in \cite{Carrillo-et-al:2020} that will provide the lacking discrete-time energy conservation. 

To integrate the differential equation \eqref{eq:fk-vdot}, I propose the following scheme
\begin{align}
    \label{eq:fk-simple-integrator}
    \frac{\bm{v}_p^{n+1}-\bm{v}_p^{n}}{\Delta t}=\frac{\nu}{m}\sum_{\bar{p}}w_{\bar{p}} \mathbf{1}(p,\bar{p})\mathbb{Q}(\bm{v}_p^{n+1/2}-\bm{v}_{\bar{p}}^{n+1/2})\cdot\bm{\Gamma}(S^n_\epsilon,p,\bar{p}),
\end{align}
where the midpoint velocity is defined according to
\begin{align}
    \bm{v}_p^{n+1/2}=\frac{\bm{v}_p^{n+1}+\bm{v}_p^{n}}{2},
\end{align}
and the notation $A^n=A(\bm{X},\bm{V}^n;W)$ refers to the function $A$ being evaluated with respect to $\bm{v}_p^n$. This scheme will conserve the kinetic energy for
\begin{align}
    \frac{K^{n+1}-K^n}{\Delta t}&=\sum_pw_pm\bm{v}_p^{n+1/2}\cdot\frac{\bm{v}_p^{n+1}-\bm{v}_p^n}{\Delta t}\nonumber\\
    &=\nu\sum_{p,\bar{p}}w_pw_{\bar{p}} \mathbf{1}(p,\bar{p})\bm{v}_p^{n+1/2}\cdot\mathbb{Q}(\bm{v}_p^{n+1/2}-\bm{v}_{\bar{p}}^{n+1/2})\cdot\bm{\Gamma}(S^n_\epsilon,p,\bar{p})\nonumber\\
    &=\frac{\nu}{2}\sum_{p,\bar{p}}w_pw_{\bar{p}}\mathbf{1}(p,\bar{p})( \bm{v}_p^{n+1/2}-\bm{v}_{\bar{p}}^{n+1/2})\cdot\mathbb{Q}(\bm{v}_p^{n+1/2}-\bm{v}_{\bar{p}}^{n+1/2})\cdot\bm{\Gamma}(S^n_\epsilon,p,\bar{p})\nonumber\\
    &=0,
\end{align}
owing the the projection property of the matrix $\mathbb{Q}$. The scheme also preserves the discrete-time kinetic momentum for 
\begin{align}
    \frac{\bm{P}^{n+1}-\bm{P}^n}{\Delta t}&=\sum_pw_pm\frac{\bm{v}_p^{n+1}-\bm{v}_p^n}{\Delta t}\nonumber\\
    &=\nu\sum_{p,\bar{p}}w_pw_{\bar{p}}\mathbf{1}(p,\bar{p}) \mathbb{I}\cdot\mathbb{Q}(\bm{v}_p^{n+1/2}-\bm{v}_{\bar{p}}^{n+1/2})\cdot\bm{\Gamma}(S^n_\epsilon,p,\bar{p})\nonumber\\
    &=\frac{\nu}{2}\sum_{p,\bar{p}}w_pw_{\bar{p}}\mathbf{1}(p,\bar{p}) (\mathbb{I}-\mathbb{I})\cdot\mathbb{Q}(\bm{v}_p^{n+1/2}-\bm{v}_{\bar{p}}^{n+1/2})\cdot\bm{\Gamma}(S^n_\epsilon,p,\bar{p})\nonumber\\
    &=0.
\end{align}
    

\subsection{A discrete-gradient integrator}
Finally, to guarantee not only discrete-time energy and momentum conservation but also discrete-time entropy dissipation, I turn the attention to the so-called discrete gradients. While many such constructions exist \cite{Harten_Lax_Leer:1983,Gonzalez:1996}, I will focus on one specific, intuitive example.

Consider the mean-value or average gradient 
\begin{align}
    \int_0^1\frac{\partial A}{\partial \bm{v}_p}(\bm{X},(1-\xi)\bm{V}^n+\xi\bm{V}^{n+1};W)d\xi\equiv \overline{\frac{\partial A}{\partial \bm{v}_p}}\Big|_{n}^{n+1}\, ,
\end{align}
which, owing to the fundamental theorem of calculus, naturally has the following property
\begin{align}
    \sum_{p}(\bm{v}_p^{n+1}-\bm{v}_p^n)\cdot\overline{\frac{\partial A}{\partial \bm{v}_p}}\Big|_{n}^{n+1}=A^{n+1}-A^n.
\end{align}
This mean-value property, though, is not unique to the mean-value or average gradient but is, instead, the defining feature of all of the so-called discrete gradients.  

The importance of the mean-value property of discrete gradients is in that it can be exploited to construct a rather convenient integrator. Specifically, I choose an integrator
\begin{align}
    \label{eq:discrete-vdot-final}
    \frac{\bm{v}_p^{n+1}-\bm{v}_p^n}{\Delta t}=\frac{\nu}{m}\sum_{\bar{p}}w_{\bar{p}}\mathbf{1}(p,\bar{p})\mathbb{Q}(\overline{\bm{\Gamma}_{n}^{n+1}(K,p,\bar{p})})\cdot \overline{\bm{\Gamma}_{n}^{n+1}(S_\epsilon,p,\bar{p})} 
\end{align}
where the operator $\overline{\bm{\Gamma}_{n}^{n+1}(A,p,\bar{p})}$ is the discrete-gradient analog of $\bm{\Gamma}(A,p,\bar{p})$, namely
\begin{align}
    \overline{\bm{\Gamma}_{n}^{n+1}(A,p,\bar{p})}=\frac{1}{mw_p}\overline{\frac{\partial A}{\partial \bm{v}_p}}\Big|_{n}^{n+1}-\frac{1}{mw_{\bar{p}}}\overline{\frac{\partial A}{\partial \bm{v}_{\bar{p}}}}\Big|_{n}^{n+1}.
\end{align}
Consequently, owing to the mean-value property, the collisional discrete-time evolution of an arbitrary function $A$ is given by 
\begin{align}
    \label{eq:fk-discrete-ode-final}
    \frac{A^{n+1}-A^n}{\Delta t}    &=\frac{1}{2}\sum_{p,\bar{p}}\overline{\bm{\Gamma}_{n}^{n+1}(A,p,\bar{p})}\cdot \overline{W_{n}^{n+1}(p,\bar{p})}\cdot \overline{\bm{\Gamma}_{n}^{n+1}(S_\epsilon,p,\bar{p})},
\end{align}
where the matrix $\overline{W_{n}^{n+1}(p,\bar{p})}$ is defined according to
\begin{align}
    \overline{W_{n}^{n+1}(p,\bar{p})}=\nu w_pw_{\bar{p}}\mathbf{1}(p,\bar{p})\mathbb{Q}(\overline{\bm{\Gamma}_{n}^{n+1}(K,p,\bar{p})}).
\end{align}
Obviously, the expression \eqref{eq:fk-discrete-ode-final} is a discrete-time approximation of the continuous-time relation \eqref{eq:fk-ode}.

Using \eqref{eq:fk-discrete-ode-final}, it is then straightforward to demonstrate the desired properties of the integrator \eqref{eq:discrete-vdot-final}. The discrete-time collisional rate-of-change of kinetic energy vanishes for
\begin{align}
    \frac{K^{n+1}-K^n}{\Delta t}
    &=\frac{\nu}{2}\sum_{p,\bar{p}}\overline{\bm{\Gamma}_{n}^{n+1}(K,p,\bar{p})}\cdot \overline{W_{n}^{n+1}(p,\bar{p})}\cdot \overline{\bm{\Gamma}_{n}^{n+1}(S_\epsilon,p,\bar{p})}\nonumber\\
    &=0,
\end{align}
owing again to the projection property
    $\overline{\bm{\Gamma}_{n}^{n+1}(K,p,\bar{p})}\cdot \mathbb{Q}(\overline{\bm{\Gamma}_{n}^{n+1}(K,p,\bar{p})})=0$.
Similarly, the discrete-time collisional rate-of-change of kinetic momentum vanishes
\begin{align}
    \frac{\bm{P}^{n+1}-\bm{P}^n}{\Delta t}
    &=\frac{\nu}{2}\sum_{p,\bar{p}}\overline{\bm{\Gamma}_{n}^{n+1}(\bm{P},p,\bar{p})}\cdot \overline{W_{n}^{n+1}(p,\bar{p})}\cdot \overline{\bm{\Gamma}_{n}^{n+1}(S_\epsilon,p,\bar{p})}\nonumber\\
    &=0,
\end{align}
owing to the result 
    $\overline{\bm{\Gamma}_{n}^{n+1}(\bm{P},p,\bar{p})}
    =\mathbb{I}-\mathbb{I}=0$.
Finally, the discrete-time entropy is dissipated
\begin{align}
    \frac{S^{n+1}_\epsilon-S^n_\epsilon}{\Delta t}
    &=\frac{\nu}{2}\sum_{p,\bar{p}}\overline{\bm{\Gamma}_{n}^{n+1}(S_\epsilon,p,\bar{p})}\cdot \overline{W_{n}^{n+1}(p,\bar{p})}\cdot \overline{\bm{\Gamma}_{n}^{n+1}(S_\epsilon,p,\bar{p})}\nonumber\\
    &\ge 0,
\end{align}
which follows from the fact that the matrix $\overline{W_{n}^{n+1}(p,\bar{p})}$ is positive semidefinite. Effectively, all of the continuous-time properties of the marker-particle discretization have now been successfully translated to discrete time.

At this point, it is possible to \revision{reflect back on} how the specific, apparently out-of-the-blue choice of $\mathbb{Q}(\bm{v}_p^{n+1/2}-\bm{v}_{\bar{p}}^{n+1/2})$ in the simple integrator \eqref{eq:fk-simple-integrator} came to be. An explicit calculation with the average discrete gradient produces
\begin{align}
    \frac{1}{mw_p}\overline{\frac{\partial K}{\partial \bm{v}_p}}\Big|_{n}^{n+1}=\int_0^1\left((1-\xi)\bm{v}_p^n+\xi \bm{v}_p^{n+1}\right)d\xi=\bm{v}_p^{n+1/2},
\end{align}
and consequently 
    $\overline{\bm{\Gamma}_{n}^{n+1}(K,p,\bar{p})}=\bm{v}_p^{n+1/2}-\bm{v}_{\bar{p}}^{n+1/2}$. 
\revision{This means that the out-of-the-blue choice of $\mathbb{Q}(\bm{v}_p^{n+1/2}-\bm{v}_{\bar{p}}^{n+1/2})$ in \eqref{eq:fk-simple-integrator} is in fact a convenient special case of the more general $\mathbb{Q}(\overline{\bm{\Gamma}_{n}^{n+1}(K,p,\bar{p})})$ in \eqref{eq:discrete-vdot-final}. While this small detail might appear irrelevant, it is only with the help of the systematic formulation involving discrete gradients that discrete-time-energy conservation can be achieved also for the more complicated gyrokinetic collision operator.} 

As a note, I add that the average or mean-value discrete gradient is not the only choice that succeeds in achieving the desired properties. In fact, if using the integrator \eqref{eq:discrete-vdot-final}, any discrete-gradient construction would be sufficient for the only structural property, that doesn't rely either on the projection property or the positive semidefiniteness of the matrix $\mathbb{Q}$, is the condition
\begin{align}
    \overline{\bm{\Gamma}_{n}^{n+1}(\bm{P},p,\bar{p})}=0.
\end{align}
Since $\bm{P}$ is linear in the particle degrees of freedom, and all discrete gradients are exact for linear functions, this property will be satisfied by any discrete-gradient construction. One such example is the so-called mid-point discrete gradient by Gonzalez \cite{Gonzalez:1996}. In numerical applications, this mid-point discrete gradient is likely also more useful than the average discrete gradient for it doesn't require any integrations, only evaluations of the function and its gradient.

\section{The electrostatic gyrokinetic Landau operator}\label{sec:gk-landau}
Next, I will apply the ideas from the previous section and construct a discrete-time energy-preserving and entropy-dissipating integrator for the electrostatic gyrokinetic Landau operator. Before jumping to the metric bracket accounting for collisions, some clarification is necessary regarding the notation and context. The following is adapted from \cite{Burby_2015_collisions,Hirvijoki_Burby_metriplectic_gk}.

\subsection{The electrostatic gyrokinetic model}
The collisional electrostatic gyrokinetic system consists of a dynamic kinetic equation and a static Gauss' law for the electric field
\begin{align}
\frac{\partial F_s}{\partial t}+\{F_s,H^{\text{gy}}_s\}^{\text{gc}}_s&=\sum_{\bar{s}}C^{\text{gy}}_{s\bar{s}}(F_s,F_{\bar{s}}),\\
\nabla\cdot\bm{E}&=4\pi(\rho_{\text{\text{gy}}}-\nabla\cdot\bm{P}).
\end{align}
In this model, the guiding-center Poisson bracket is given by
\begin{align}
\{F,G\}^{\text{gc}}=&\frac{e}{mc}\left(\frac{\partial F}{\partial \theta}\frac{\partial G}{\partial \mu}-\frac{\partial F}{\partial\mu}\frac{\partial G}{\partial \theta}\right)-\frac{c\bm{b}}{eB_{\parallel}^{*}}\cdot\left(\nabla^{*}F\times\nabla^{*}G\right)\nonumber\\
&+\frac{\bm{B}^{*}}{mB_{\parallel}^{*}}\cdot\left(\nabla^{*}F\frac{\partial G}{\partial v_{\parallel}}-\frac{\partial F}{\partial v_{\parallel}}\nabla^{*}G\right),
\end{align}
and $H^{\text{gy}}=K^{\text{gy}}+e\varphi$ is the gyrocenter Hamiltonian. The polarization density is given by 
\begin{align}
\bm{P}=-\delta\mathcal{K}/\delta\bm{E},\qquad \mathcal{K}=\sum_s \int K^{\text{\text{gy}}}_s\,F_s\,d\bm{z}_s^{\text{\text{gc}}},
\end{align} 
and the function $K^{\text{\text{gy}}}$, appearing in the Hamiltonian $H^{\text{gy}}=K^{\text{gy}}+e\varphi$, is the gyrocenter kinetic energy, which may be written entirely in terms of the electric field as
\begin{align}
K^{\text{\text{gy}}}=&\frac{1}{2}m v_\parallel^2+\mu |\bm{B}|
-e\langle \llbracket\bm{\rho}_0\cdot\bm{E}(\bm{X}+\epsilon\bm{\rho}_0)\rrbracket\rangle\nonumber\\
&-\frac{e^2}{2\mu|\bm{B}|}\langle\llbracket \widetilde{\bm{\rho}_0\cdot\bm{E}}(\bm{X}+\epsilon\bm{\rho}_0)\widetilde{\bm{\rho}_0\cdot\bm{E}}(\bm{X}+\bm{\rho}_0)\rrbracket\rangle\nonumber\\
&-\frac{e^2}{2m \omega_c^2}\bm{b}\cdot\langle\tilde{\bm{E}}(\bm{X}+\bm{\rho}_0)\times I\tilde{\bm{E}}(\bm{X}+\bm{\rho}_0)\rangle.
\end{align}
Here $\langle\cdot\rangle_{s}=(2\pi)^{-1}\int_0^{2\pi}\cdot\,d\theta_{s}$ denotes the average with respect to the species-$s$ gyroangle, tildes denote the fluctuating part of a gyroangle-dependent quantity, $I=\partial_\theta^{-1}$ is the gyroangle antiderivative,  $\llbracket\cdot\rrbracket=\int_0^1\cdot\,d\epsilon$, and $\bm{\rho}_0$ is the zero'th order (gyroangle-dependent) gyroradius vector. 

The collision operator for this system was derived in \cite{Burby_2015_collisions} and is given by 
\begin{align}
    \label{eq:gk-collision-operator}
     C^{\text{gy}}_{s\bar{s}}(F_s,F_{\bar{s}})(\bm{z})=-\nu_{s\bar{s}}\left\langle\left\{y^i_s(\bm{z}),\gamma_{s\bar{s}}^i(\bm{z})\right\}_{s}^{\text{gc}}\right\rangle_{s},
\end{align}
where the three-component vector $\bm{\gamma}_{s\bar{s}}(\bm{z})$ can be written as
\begin{align}
     \bm{\gamma}_{s\bar{s}}(\bm{z})=\int \delta(\bm{y}_{s}(\bm{z})-\bm{y}_{\bar{s}}(\bm{\bar{z}}))F_s(\bm{z})F_{\bar{s}}(\bm{\bar{z}})\mathbb{Q}(\bm{u}^{\text{gy}}_{s\bar{s}}(\bm{z},\bm{\bar{z}}))\cdot\bm{\Gamma}^{\text{gy}}_{s\bar{s}}({\cal S}^{\text{gy}},\bm{z},\bm{\bar{z}})d\bm{\bar{z}}^{\text{gc}}.
\end{align}
The expression I have written for $\bm{\gamma}_{s\bar{s}}$ is the same as in \cite{Burby_2015_collisions} but slightly re-organized, to be directly expressible in terms of the entropy functional
\begin{align}
    \mathcal{S}^{\text{gy}}=-\sum_s \int F_s(\bm{z})\ln F_s(\bm{z})d\bm{z}^{\text{gc}}_s.
\end{align}
In the expression for $\bm{\gamma}_{s\bar{s}}$, the vector $\bm{\Gamma}^{\text{gy}}_{s\bar{s}}({\cal A},\bm{z},\bm{\bar{z}})$ is defined according to
\begin{align}
    \bm{\Gamma}^{\text{gy}}_{s\bar{s}}({\cal A},\bm{z},\bm{\bar{z}})&=\left\{\bm{y}_s,\frac{\delta {\cal A}}{\delta F_s}\right\}^{\text{gc}}_{s}(\bm{z})-\left\{\bm{y}_{\bar{s}},\frac{\delta {\cal A}}{\delta F_{\bar{s}}}\right\}^{\text{gc}}_{\bar{s}}(\bm{\bar{z}}),
\end{align}
the relative particle velocity is computed as
\begin{align}
    \bm{u}^{\text{gy}}_{s\bar{s}}(\bm{z},\bm{\bar{z}})&=\{\bm{y}_s,H^{\text{gy}}_s\}^{\text{gc}}_s(\bm{z})-\{\bm{y}_{\bar{s}},H^{\text{gy}}_s\}^{\text{gc}}_{\bar{s}}(\bm{\bar{z}}),
\end{align}and the particle position in terms of the gyrocenter variables is given by
\begin{align}
    \bm{y}_{s}(\bm{z})=\bm{x}+\bm{\rho}_{0s}(\bm{x},\mu,\theta).
\end{align}

This collision operator is unique from all other gyrokinetic collision operators considered in literature for it conserves the energy
\begin{align}
    \mathcal{E}^{\text{gy}}=\sum_s\int H_s^{\text{gy}}F_sd\bm{z}^{\text{gc}}_s,
\end{align}
in the sense that, in collisions between any two species $s$ and $\bar{s}$, one finds the relation
\begin{align}
    \label{eq:gk-energy-conservation}
    \int H^{\text{gy}}_sC^{\text{gy}}_{s\bar{s}}(F_s,F_{\bar{s}})d\bm{z}^{\text{gc}}_s+\int H^{\text{gy}}_{\bar{s}}C^{\text{gy}}_{\bar{s}s}(F_{\bar{s}},F_s)d\bm{z}^{\text{gc}}_{\bar{s}}=0.
\end{align}
Similarly, in case of an axially symmetric magnetic field $\bm{B}$, also the toroidal canononical momentum 
\begin{align}
    \label{eq:gk-toroidal-momentum}
    \mathcal{P}_{\phi}=\sum_s\int p_{\phi s}F_sd\bm{z}^\text{gc}_s,
\end{align}
is conserved for the collision operator satisfies also the relation
\begin{align}
    \label{eq:gk-momentum-conservation}
    \int p_{\phi s}C^{\text{gy}}_{s\bar{s}}(F_s,F_{\bar{s}})d\bm{z}^{\text{gc}}_s+\int p_{\phi \bar{s}}C^{\text{gy}}_{\bar{s}s}(F_{\bar{s}},F_s)d\bm{z}^{\text{gc}}_{\bar{s}}=0.
\end{align}
The details of these calculations can be found in \cite{Burby_2015_collisions}. The operator also dissipates the entropy functional and has a Maxwellian as the equilibrium state.

\subsection{Metric formulation of the collision operator}
A detailed account on the metriplectic structure of the electrostatic gyrokinetic model described in the previous section can be found in \cite{Hirvijoki_Burby_metriplectic_gk}. Here I briefly outline how the metric formulation of the collision operator emerges.

For any three functions $F$, $G$, and $H$, a Poisson bracket satisfies the identity
\begin{align}
    \{F,G\}H=\{F,GH\}-\{F,H\}G.
\end{align}
Inspired by the conservation relations \eqref{eq:gk-energy-conservation} and \eqref{eq:gk-momentum-conservation}, I then investigate a more general case of collisional evolution of some species-dependent functions $G_s$. I multiply the collision operator \eqref{eq:gk-collision-operator} with a gyroangle independent test function $G_s$, apply the Poisson-bracket identity, and integrate over the entire phase-space. This provides 
\begin{align}
    \label{eq:gk-collisional-evolution}
    &\sum_{s,\bar{s}}\int G_s C_{s\bar{s}}^{\text{gy}}(F_s,F_{\bar{s}})d\bm{z}^{\text{gc}}_s\nonumber\\
    &=\sum_{s,\bar{s}}\iint \{\bm{y}_s,G_s\}^{\text{gc}}_s(\bm{z})\cdot\mathbb{W}^{\text{gy}}_{s\bar{s}}(\bm{z},\bm{\bar{z}})\cdot\bm{\Gamma}^{\text{gy}}_{s\bar{s}}({\cal S}^{\text{gy}},\bm{z},\bm{\bar{z}})d\bm{z}^{\text{gc}}_{\bar{s}}d\bm{z}^{\text{gc}}_s\nonumber\\
    &=\sum_{s,\bar{s}}\frac{1}{2}\iint \left(\{\bm{y}_s,G_s\}^{\text{gc}}_s(\bm{z})-\{\bm{y}_{\bar{s}},G_{\bar{s}}\}^{\text{gc}}_{\bar{s}}(\bm{\bar{z}})\right)\cdot\mathbb{W}^{\text{gy}}_{s\bar{s}}(\bm{z},\bm{\bar{z}})\cdot\bm{\Gamma}^{\text{gy}}_{s\bar{s}}({\cal S}^{\text{gy}},\bm{z},\bm{\bar{z}})d\bm{z}^{\text{gc}}_{\bar{s}}d\bm{z}^{\text{gc}}_s
\end{align}
where the symmetric, positive semi-definite matrix $\mathbb{W}_{s\bar{s}}^{\text{gy}}(\bm{z},\bm{\bar{z}})$ is
\begin{align}
    \label{eq:gy-W-matrix}
    \mathbb{W}^{\text{gy}}_{s\bar{s}}(\bm{z},\bm{\bar{z}})&=\nu_{s\bar{s}}\delta(\bm{y}_{s}(\bm{z})-\bm{y}_{\bar{s}}(\bm{\bar{z}}))F_s(\bm{z})F_{\bar{s}}(\bm{\bar{z}})\mathbb{Q}(\bm{u}^{\text{gy}}_{s\bar{s}}(\bm{z},\bm{\bar{z}})),
\end{align}
and the last line follows from the antisymmetry of the vector $\bm{\Gamma}^{\text{gy}}_{s\bar{s}}({\cal S},\bm{z},\bm{\bar{z}})$ with respect to exchanging both the coordinates and the species labels. 

If I now consider the function $G_s$ to be expressed as a functional derivative
\begin{align}
    \mathcal{G}=\sum_s\int G_sF_sd\bm{z}^{\text{gc}}_s \qquad \implies  \qquad \frac{\delta G}{\delta F_s}=G_s,
\end{align}
I can identify $\bm{\Gamma}^{\text{gy}}_{s\bar{s}}({\cal G},\bm{z},\bm{\bar{z}})$ in \eqref{eq:gk-collisional-evolution} and define a  symmetric, positive semidefinite bracket
\begin{align}
    \label{eq:gk-bracket}
    ({\cal A},{\cal B})=\sum_{s\bar{s}}\frac{1}{2}\iint\bm{\Gamma}^{\text{gy}}_{s\bar{s}}({\cal A},\bm{z},\bm{\bar{z}})\cdot\mathbb{W}^{\text{gy}}_{s\bar{s}}(\bm{z},\bm{\bar{z}})\cdot\bm{\Gamma}^{\text{gy}}_{s\bar{s}}({\cal B},\bm{z},\bm{\bar{z}})d\bm{z}^{\text{gc}}_{\bar{s}}d\bm{z}^{\text{gc}}_s.
\end{align}
In terms of this bracket, the collisional evolution of arbirtrary functionals can now be generatlized to the functional equation
\begin{align}
    \frac{d\mathcal{A}}{dt}\Big|_{\text{coll}}=(\mathcal{A},\mathcal{S}^{\text{gy}}).
\end{align}
The properties of this bracket have been discussed in detail in \cite{Hirvijoki_Burby_metriplectic_gk}. In brief, it has the system Hamiltonian functional 
\begin{align}
    \mathcal{H}^{\text{gy}}=\sum_s\int H_s^{\text{gy}}F_sd\bm{z}_s^{\text{gc}}-\frac{1}{8\pi}\int |\bm{E}|^2 d\bm{x},
\end{align}
as an invariant in the sense of $(\mathcal{H},\mathcal{A})=0$ and, in case of an axially symmetric magnetic background, also the total toroidal angular momentum \eqref{eq:gk-toroidal-momentum}, in the sense of $(\mathcal{P}_{\phi},\mathcal{A})=0$, with respect to arbitrary functionals $\mathcal{A}$. Finally, as the bracket is positive semidefinite by construction, the entropy will be dissipated.

For the rest of this paper, I will again consider only the single-species case to avoid clutter of indices. Consequently the species indices $s,\bar{s}$ and the associated summations are dropped.

\subsection{Discretization of the metric bracket}
Consider a discrete distribution function such that one can formally express the phase-space density as
\begin{align}
    F_h(\bm{z},t)d\bm{z}^{\text{gc}}=\sum_p w_p\delta(\bm{x}-\bm{x}_p(t))\delta(v_\parallel-v_{\parallel p}(t))\delta(\mu-\mu_p(t))d\bm{x}dv_\parallel d\mu d\theta.
\end{align}
Note that $d\bm{z}^{\text{gc}}$ is not the bare differential volume but contains also the phase-space Jacobian. Similarly as previously, the argument of scalar invariance $\mathcal{A}[F_h]=A(\bm{Z})$, with $\bm{Z}=\{\bm{z}_p\}_p$ denoting the collection of particle degrees of freedom and $W=\{w_p\}_p$ the collection of weights, provides the discretization of the functional derivatives
\begin{align}
    \frac{\partial }{\partial z^\alpha}\frac{\delta {\cal A}}{\delta F}\Big|_{\bm{z}_p}=\frac{1}{w_p}\frac{\partial A(\bm{Z};W)}{\partial z_p^\alpha}.
\end{align}
Approximating the delta function in the matrix $\mathbb{W}^{\text{gy}}$ \eqref{eq:gy-W-matrix} with an indicator function that evaluates to one or zero depending on whether two points are within the same collisional cell, I substitute everything into the single-species version of the metric bracket \eqref{eq:gk-bracket}. 

The result is a finite-dimensional metric bracket operating on regular functions
\begin{align}
    \label{eq:gk-bracket-discrete}
    (A,B)_h=\frac{1}{2}\sum_{p,\bar{p}}\langle\langle\bm{\Gamma}^{\text{gy}}(A,p,\bar{p})\cdot\mathbb{W}^{\text{gy}}(p,\bar{p})\cdot\bm{\Gamma}^{\text{gy}}(B,p,\bar{p})\rangle_{p}\rangle_{\bar{p}},
\end{align}
where the vector $\bm{\Gamma}^{\text{gy}}(A,p,\bar{p})$ is now given by
\begin{align}
    \bm{\Gamma}^{\text{gy}}(A,p,\bar{p})=\{\bm{y},z^\alpha\}^{\text{gc}}(\bm{z}_p)\frac{1}{w_p}\frac{\partial A}{\partial z_p^\alpha}-\{\bm{y},z^\alpha\}^{\text{gc}}(\bm{z}_{\bar{p}})\frac{1}{w_{\bar{p}}}\frac{\partial A}{\partial z_{\bar{p}}^\alpha},
\end{align}
the matrix $\mathbb{W}^{\text{gy}}(p,\bar{p})$ is given by
\begin{align}
    \mathbb{W}^{\text{gy}}(p,\bar{p})=\nu w_pw_{\bar{p}}\mathbf{1}^{\text{gy}}(p,\bar{p})\mathbb{Q}(\bm{u}^{\text{gy}}(\bm{z}_p,\bm{z}_{\bar{p}})),
\end{align}
and the indicator function $\mathbf{1}^{\text{gy}}(p,\bar{p})$ checks the for the actual particle positions $\bm{y}(\bm{z}_p)$ and $\bm{y}(\bm{z}_{\bar{p}})$ instead of just the gyrocenter positions $\bm{x}_p$ and $\bm{x}_{\bar{p}}$. The bracket $\langle \cdot \rangle_p$ refers to gyroaverage with respect to the gyroangle of the particle $p$, an operation that is computed by sampling a given number of angles and computing the average. In practice, even a single sample value could be acceptable for this will not affect the energy conservation nor entropy dissipation. 

Given a regularized entropy functional $S_\epsilon^{\text{gy}}$, computed via
\begin{align}
    S^{\text{gy}}_\epsilon(\bm{Z};W)&=\mathcal{S}^{\text{gy}}[\psi_\epsilon\ast F_h]
    =-\int \sum_pw_p\psi_\epsilon(\bm{z}-\bm{z}_p)\ln\left(\sum_{p'}w_{p'}\psi_\epsilon(\bm{z}-\bm{z}_{p'})\right)d\bm{z}^{\text{gc}},
\end{align}
the finite-dimensional bracket \eqref{eq:gk-bracket-discrete} then provides collisional evolution of functions that depend on the particle degrees of freedom 
\begin{align}
    \frac{d A}{dt}\Big|_{\text{coll}}=(A,S^{\text{gy}}_\epsilon)_h.
\end{align}
Of specific interest is the case $A=z_p^{\alpha}$, which provides the equation of motion for the particle coordinates. Exploiting the antisymmetry of the vector $\bm{\Gamma}^{\text{gy}}(A,p,\bar{p})$ with respect to exchanging $p$ and $\bar{p}$, the result can be expressed as
\begin{align}
    \frac{dz_p^\alpha}{dt}\Big|_{\text{coll}}=\nu\sum_{\bar{p}}w_{\bar{p}}\langle\langle\{\bm{y},z^\alpha\}^{\text{gc}}(\bm{z}_p)\cdot \mathbf{1}^{\text{gy}}(p,\bar{p})\mathbb{Q}(\bm{u}^{\text{gy}}(\bm{z}_p,\bm{z}_{\bar{p}}))\cdot\bm{\Gamma}^{\text{gy}}(S_\epsilon^{\text{gy}},p,\bar{p})\rangle_{p}\rangle_{\bar{p}}.
\end{align}

The finite-dimensional bracket \eqref{eq:gk-bracket-discrete} also preserves the finite-dimensional version of the energy
\begin{align}
    E(\bm{Z};W)=\mathcal{E}^{\text{gy}}[F_h]=\sum_pw_pH^{\text{gy}}(\bm{z}_p),
\end{align}
for the simple reason that the discrete energy is an invariant of the discrete bracket 
\begin{align}
    (E,A)_h=\frac{1}{2}\sum_{p,\bar{p}}\langle\langle\bm{\Gamma}^{\text{gy}}(E,p,\bar{p})\cdot\mathbb{W}^{\text{gy}}(p,\bar{p})\cdot\bm{\Gamma}^{\text{gy}}(A,p,\bar{p})\rangle_{p}\rangle_{\bar{p}}=0.
\end{align}
This is easy to verify after realizing that
\begin{align}
    \bm{\Gamma}^{\text{gy}}(E,p,\bar{p})=\bm{u}^{\text{gy}}(\bm{z}_p,\bm{z}_{\bar{p}}),
\end{align}
and consequently $\bm{\Gamma}^{\text{gy}}(E,p,\bar{p})\cdot \mathbb{W}^{\text{gy}}(p,\bar{p})=0$. As the discrete bracket is also positive semidefinite by construction, the regularized entropy is dissipated
\begin{align}
    \frac{dS_\epsilon^{\text{gy}}}{dt}\Big|_{\text{coll}}=(S_\epsilon^{\text{gy}},S_\epsilon^{\text{gy}})_h\ge 0.
\end{align}

Unlike in the infinite-dimensional case, the discrete bracket unfortunately does not \revision{precisely} conserve the discrete version of the toroidal canonical momentum
\begin{align}
    P_\phi=\mathcal{P}_{\phi}[F_h]=\sum_sw_pp_{\phi}(\bm{z}_p).
\end{align}
\revision{The reason for this lies in the particles' participating in collisions within one collision cell having disctinct locations in spatial coordines and the conservation of toroidal angular momentum in the infinite-dimensional case relying on strict localization of the collisions. Even if a direct computation in an axially symmetric magnetic background still provides
\begin{align}
    \bm{\Gamma}^{\text{gy}}(P_\phi,p,\bar{p})=\hat{z}\times(\bm{y}(\bm{z}_p)-\bm{y}(\bm{z}_{\bar{p}})),
\end{align}
with $\hat{z}$ the unit vector along the rotational symmetry axis of the magnetic field, which is essentially the same expression as in the infinite-dimensional case, there is no delta function and integration left to kill the bracket $(P_\phi,A)_h$ in the same way they exist in the infinite-dimensional case (see \cite{Burby_2015_collisions,Hirvijoki_Burby_metriplectic_gk} for details). This is an unfortunate but unavoidable consequence of using marker particles. Exactly quantifying this error and its significance would require performing electrostatic gyrokinetic full-$f$ particle-in-cell simulations and is beyond the scope of the present study. Nevertheless, it is useful to understand that the error should be controllable by especially the toroidal size of the collision cell.}

\subsection{A discrete-gradient integrator}
In the case of the standard Landau operator, \revision{it was reasonably straightforward to identify a simple energy-preserving integrator by exploiting the identity $|\bm{v}^{n+1}|^2-|\bm{v}^{n}|^2=(\bm{v}^{n+1}+\bm{v}^{n})\cdot(\bm{v}^{n+1}-\bm{v}^n)$.} In dealing with the gyrokinetic version, no such luxury exists. The energy function is far more complicated and, for now at least, it seems that the concept of discrete gradients is the only approach forward.

Guided by the analysis of the regular Landau operator case, I define the operator
\begin{align}
    \overline{\bm{\Gamma}^{\text{gy}}_{n,n+1}(A,p,\bar{p})} =\{\bm{y},z^\alpha\}^{\text{gc}}(\bm{z}_p^{n+1/2})\frac{1}{w_p}\overline{\frac{\partial A}{\partial z^\alpha_p}}\Big|_n^{n+1}-\{\bm{y},z^\alpha\}^{\text{gc}}(\bm{z}_{\bar{p}}^{n+1/2})\frac{1}{w_{\bar{p}}}\overline{\frac{\partial A}{\partial z^\alpha_{\bar{p}}}}\Big|_n^{n+1},
\end{align}
where the discrete gradient could again be computed, e.g., as the mean-value gradient
\begin{align}
    \overline{\frac{\partial A}{\partial z^\alpha_p}}\Big|_n^{n+1}=\int_0^1\frac{\partial A}{\partial z^\alpha_p}((1-\xi)\bm{Z}^n+\xi \bm{Z}^{n+1})d\xi.
\end{align}
Using the above choice and the notation $A^n=A(\bm{Z}^n;W)$, the integrator accounting for the collisional discrete-time evolution of functions $A$ can be expressed as 
\begin{align}
    \frac{A^{n+1}-A^n}{\Delta t}
    &=\frac{1}{2}\sum_{p,\bar{p}}\langle\langle\overline{\bm{\Gamma}_{n,n+1}^{\text{gy}}(A,p,\bar{p})}\cdot\overline{\mathbb{W}^{\text{gy}}_{n,n+1}(p,\bar{p})}\cdot \overline{\bm{\Gamma}_{n,n+1}^{\text{gy}}(S_\epsilon^{\text{gy}},p,\bar{p})}\rangle_p\rangle_{\bar{p}}.
\end{align}
The matrix $\overline{W^{\text{gy}}_{n,n+1}(p,\bar{p})}$ in the expression above is defined as
\begin{align}
    \overline{W^{\text{gy}}_{n,n+1}(p,\bar{p})}=\nu w_pw_{\bar{p}}\mathbf{1}^{\text{gy}}(p,\bar{p})\mathbb{Q}(\overline{\bm{\Gamma}_{n,n+1}^{\text{gy}}(E,p,\bar{p})}).
\end{align}

The specific choice of $\mathbb{Q}(\overline{\bm{\Gamma}_{n,n+1}^{\text{gy}}(E,p,\bar{p})})$ now guarantees that the discrete-time energy is conserved in the collisions
\begin{align}
    \frac{E^{n+1}-E^n}{\Delta t}
    &=\frac{1}{2}\sum_{p,\bar{p}}\langle\langle\overline{\bm{\Gamma}_{n,n+1}^{\text{gy}}(E,p,\bar{p})}\cdot\overline{\mathbb{W}^{\text{gy}}_{n,n+1}(p,\bar{p})}\cdot \overline{\bm{\Gamma}_{n,n+1}^{\text{gy}}(S_\epsilon^{\text{gy}},p,\bar{p})}\rangle_p\rangle_{\bar{p}}\nonumber
    \\
    &=0,
\end{align}
for $\overline{\bm{\Gamma}_{n,n+1}^{\text{gy}}(E,p,\bar{p})}\cdot\overline{\mathbb{W}^{\text{gy}}_{n,n+1}(p,\bar{p})}=0$. Similarly, entropy dissipation
\begin{align}
    \frac{S^{\text{gy},n+1}_\epsilon-S_\epsilon^{\text{gy},n}}{\Delta t}
    &=\frac{1}{2}\sum_{p,\bar{p}}\langle\langle\overline{\bm{\Gamma}_{n,n+1}^{\text{gy}}(S_\epsilon^{\text{gy}},p,\bar{p})}\cdot\overline{\mathbb{W}^{\text{gy}}_{n,n+1}(p,\bar{p})}\cdot \overline{\bm{\Gamma}_{n,n+1}^{\text{gy}}(S_\epsilon^{\text{gy}},p,\bar{p})}\rangle_p\rangle_{\bar{p}}\nonumber\\
    &\ge 0,
\end{align}
is quaranteed as the matrix $\overline{W^{\text{gy}}_{n,n+1}(p,\bar{p})}$ is positive semidefinite. 

Finally, the discrete equation of motion for the particle coordinates is found by choosing $A=z_p^{\alpha}$ which, upon using the antisymmetry of $\overline{\bm{\Gamma}^{\text{gy}}(S_h)}$ with respect to the indices $p$ and $\bar{p}$, provides
\begin{align}
    \frac{z_p^{\alpha,n+1}-z_p^{\alpha,n}}{\Delta t}=\nu\sum_{\bar{p}}w_{\bar{p}}\langle\langle\{\bm{y},z^{\alpha}\}^{\text{gc}}(\bm{z}_p^{n+1/2})\cdot\overline{\mathbb{W}^{\text{gy}}_{n,n+1}(p,\bar{p})}\cdot\overline{\bm{\Gamma}_{n,n+1}^{\text{gy}}(S_\epsilon^{\text{gy}},p,\bar{p})}\rangle_p\rangle_{\bar{p}}.
\end{align}

While the conservation of toroidal canonical angular momentum cannot be exactly recovered, the proposed discretization is nevertheless thermodynamically consistent: it preserves the energy and dissipates entropy.

\section{Relaxation of a double Maxwellian -- a demonstration}\label{sec:numerics}
Here, I will provide a simple demonstration to back up my claims regarding the momentum and energy conservation, by simulating the collisional relaxation of a double-Maxwellian distribution function in a reduced 2-D setting. Essentially, the demonstration will be a replication of the example 3 described in Sec. 4.3 of Ref. \cite{Carrillo-et-al:2020} but with a different algorithm. 

\revision{The physical parameters, $m$, $\nu$, as well as the integral of the distribution over the dimensionless velocity space, are set to one.} The initial state for the distribution is chosen as
\begin{align}\label{eq:initial_state}
    f(\bm{v},t=0)=\frac{1}{4\pi}\left[\exp\left(-\frac{(\bm{v}-\bm{u}_1)^2}{2}\right)+\exp\left(-\frac{(\bm{v}-\bm{u}_2)^2}{2}\right)\right],
\end{align}
where the peaks of the Maxwellians are $\bm{u}_1=(-2,1)$ and $\bm{u}_2=(0,-1)$. The energy and momentum of this distribution are $E=2.5$ and $\bm{P}=(-1,0)$ respectively. The radial basis function $\psi_\epsilon$ is chosen to be the Gaussian 
\begin{align}
    \psi_{\epsilon}(\bm{v})=\frac{1}{2\pi\epsilon}\exp\left(-\frac{|\bm{v}|^2}{2\epsilon}\right),
\end{align}
with $\epsilon=0.64 h^{1.98}$, the parameter $h=2 L/\sqrt{N}$, $L=10$, and the total particle number $N=60^2=3600$. The particles are initialized in a regular grid in the domain $[-L,L]\times[-L,L]$ and the weights adjusted to match the initial distribution. 
The discrete entropy functional is computed with a 2-D Gauss-Hermite quadrature, localizing a 6-by-6 mesh of quadrature points to the velocity position of each particle.

In this demonstration, I will push the particles with the implicit scheme \eqref{eq:fk-simple-integrator} \revision{using a time step of $\Delta t=1/16$.} The resulting nonlinear system I solve using fixed-point iteration with the tolerance for the iteration set to 1E-15. The program for the demonstration has been written in \verb|Python|, parallelized with \verb|CUDA| via use of the \verb|Numba| package, and the simulations have been performed on a single \verb|NVIDIA| Quaddro K2200 GPU card. The source code for the implementation is available upon request from the author (contact via the email indicated on the front page of this paper).

\begin{figure}
    \centering

    \begin{tikzpicture}[image/.style = {text width=0.45\textwidth,inner sep=0pt, outer sep=0pt},node distance = 0mm and 0mm] 
        \node [image] (frame1)
        {\includegraphics[width=\linewidth]{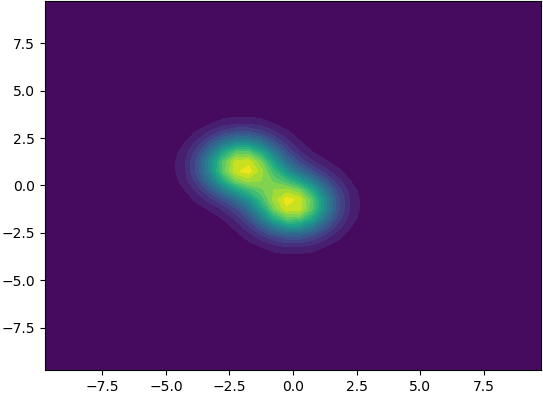}};
        
        \node [image,right=of frame1] (frame2) 
        {\includegraphics[width=\linewidth]{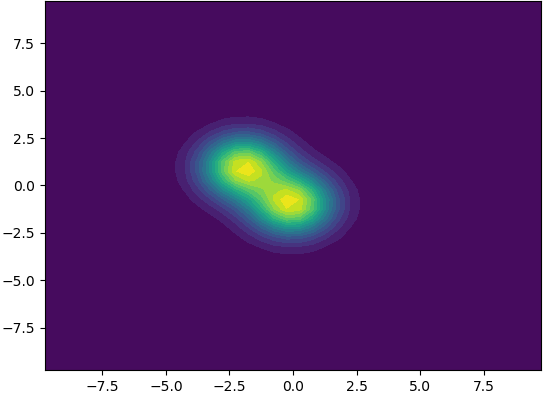}};
        
        \node[image,below=of frame1] (frame3)
        {\includegraphics[width=\linewidth]{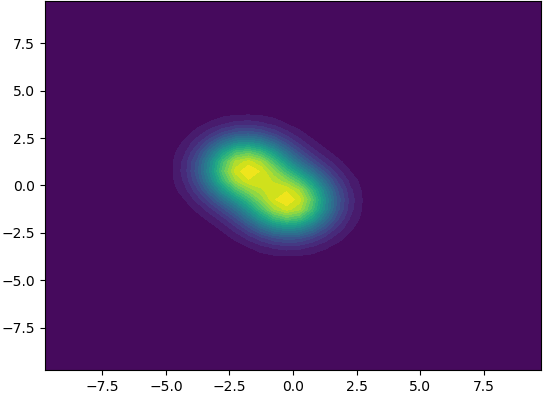}};
        
        \node[image,right=of frame3] (frame4)
        {\includegraphics[width=\linewidth]{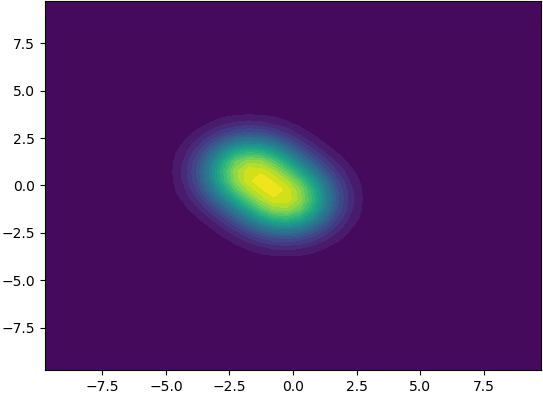}};
        
        \node[image,below =of frame3] (frame5)
        {\includegraphics[width=\linewidth]{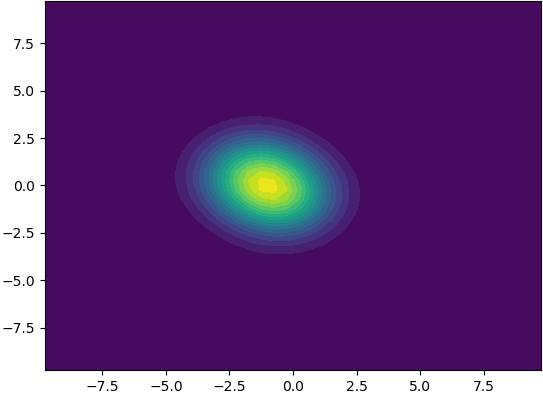}};
        
        \node[image,right=of frame5] (frame6)
        {\includegraphics[width=\linewidth]{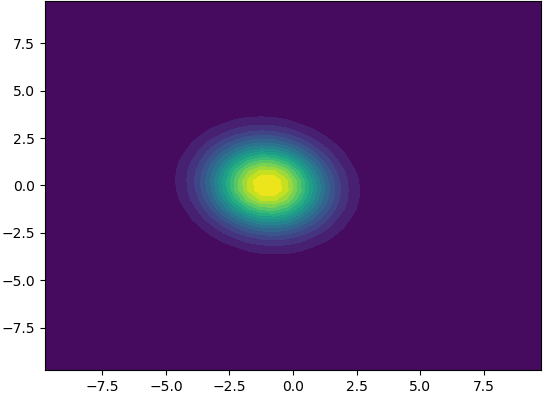}};
        
        \node[rectangle,below=of frame5] (xlabel5) {$v_1$};
        \node[rectangle,below=of frame6] (xlabel6) {$v_1$};
        \node[rectangle,left=of frame1] (ylabel1) {$v_2$};
        \node[rectangle,left=of frame3] (ylabel3) {$v_2$};
        \node[rectangle,left=of frame5] (ylabel5) {$v_2$};
        
    \end{tikzpicture}

    \caption{Collisional relaxation of a double Maxwellian \eqref{eq:initial_state}. The panels describe snapshots of the steps \#$(1,10,30,60,120,200)$ from left to right and top down. \revision{The axes in the panels refer to the velocity coordinates $(v_1,v_2)$ in the domain $[-L,L]\times[-L,L]$ and the color indicates the level sets of the distribution function from zero (deep blue) to the instantaneous maximum values (bright yellow) for optimal visual contrast.}}
    \label{fig:relaxation}
\end{figure}
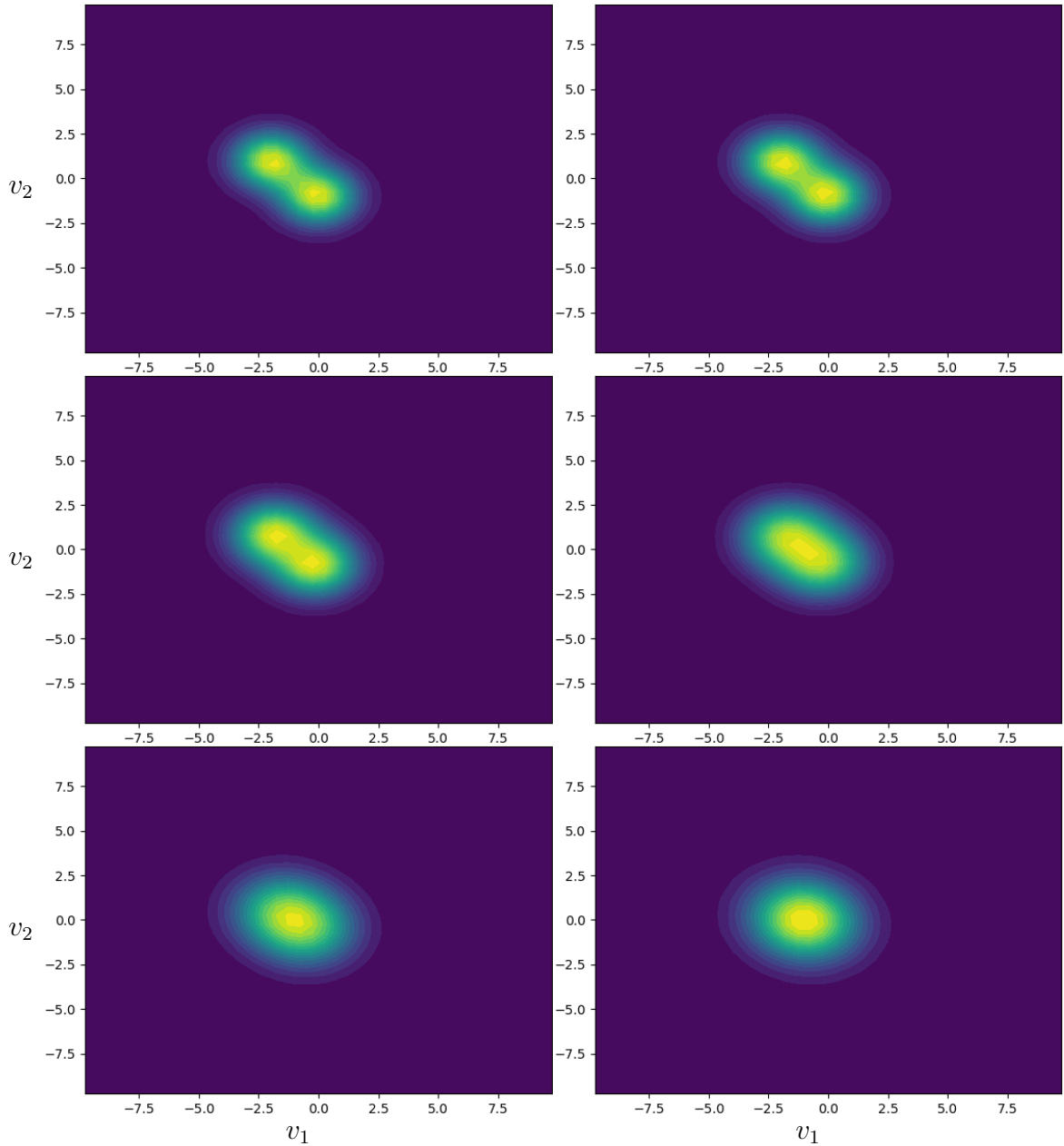

The collisional evolution of the distribution function according to the chosen implicit particle push is illustrated in Fig. \ref{fig:relaxation} by evaluating $\psi_\epsilon\ast f_h$ on a regular mesh in the domain $[-L,L]\times[-L,L]$. The panels indicate the time steps \#(1,10,30,60,120,200) from left to right and top down. The corresponding values of kinetic momentum and energy are recorded in the Table \ref{tab:conservation} and support the claims I have made. Throughout the simulation, the values for momentum and energy remain the same as the initial values up to 15 digits, demonstrating the machine-precision conservation of momentum and energy. The maximum number of fixed-point iterations per step was set to no more than 10 while the typical number of required iterations was around 4-8. A Jacobian-Free Newton solver would likely converge faster but is left for future studies and genuine HPC implementations.

\begin{table}
    \centering
    \begin{tabular}{|c|c|c|c|}
    \hline  
        Step \# & $P_1$ & $P_2$ & $E$\\
        \hline
        1 & -0.9999999999999982 & -1.8617208789871285E-16 & 2.499999999999991 \\
        10 & -0.9999999999999984 & -4.263625043299246E-16  & 2.4999999999999907 \\
        30 & -0.9999999999999981 & -1.4125799054770516E-16 & 2.500000000000011 \\
        60 & -0.9999999999999984 & -1.2262462096082616E-15 & 2.5000000000000293 \\
        120 & -0.9999999999999974 & 4.1795993749316196E-17 & 2.500000000000039 \\
        200 & -0.9999999999999982 & -4.68985202235761E-16 & 2.500000000000042 \\
        \hline
    \end{tabular}
    \caption{Conservation of momentum and energy during the collisional relaxation of a double Maxwellian. The step numbers correspond to the panels in Fig.\ref{fig:relaxation}.}
    \label{tab:conservation}
\end{table}

\section{Summary}
In this paper, I have built upon previous work and proposed new integrators for modelling Coulomb collisions in particle-in-cell codes. The first part of the paper discussed the standard Landau operator typically used in conjunction with the Vlasov-Maxwell system while the second half was devoted for the electrostatic gyrokinetic model often used in the studies of turbulent transport in tokamak and stellarator plasmas. All of the integrators were designed to preserve the first and the second law of thermodynamics, namely the preservation of energy and dissipation of entropy. In case of the standard Landau operator, also the discrete-time kinetic momentum is preserved. Finally, a numerical demonstration of the machine-precision conservation properties was provided for the standard Landau operator.

The proposed schemes are implicit but require no matrix inversion, at least if fixed-point iteration is used. Domain decomposition with respect to the spatial collision cells should be straightforward and fully analogous to the existing binary collision algorithms. Consequently, the proposed methods could potentially be implemented in existing full-$f$ particle-in-cell codes to replace binary-collision algorithms and to enable the use of arbitrary marker-particle weights. This is expected to improve the performance in simulations where the densities vary significantly over the computation domain. Furthermore, the algorithms are expected to enjoy high arithmetic intensity on modern architechture, especially on GPUs.

\section*{Acknowledgments}
I wish to thank my colleague and friend Dr. Laurent Ch\^on\'e for introducing me to the \verb|Numba| package and \verb|CUDA|, and for explaining the basics of GPU programming. Research presented in this article was supported by the Academy of Finland grant no. 315278 and contributes towards the EUROFusion enabling research project ''MAGYK: Mathematics and Algorithms for GYrokinetic and Kinetic models'' (MPG-04). Any subjective views or opinions expressed herein do not necessarily represent the views of the Academy of Finland, Aalto University, or the EUROFusion consortium.

\newpage

\bibliographystyle{unsrtnat}
\bibliography{bibfile}  

\end{document}